\documentclass[sn-mathphys-num]{sn-jnl}

\usepackage{graphicx}%
\usepackage{multirow}%
\usepackage{amsmath,amssymb,amsfonts}%
\usepackage{amsthm}%
\usepackage{mathrsfs}%
\usepackage[title]{appendix}%
\usepackage{xcolor}%
\usepackage{textcomp}%
\usepackage{manyfoot}%
\usepackage{booktabs}%
\usepackage{algorithm}%
\usepackage{algorithmicx}%
\usepackage{algpseudocode}%
\usepackage{listings}%

\begin{document}

\title[Recent advances in solar data-driven MHD simulations of the formation and evolution of CME flux ropes 
]{Recent advances in solar data-driven MHD simulations  of the formation and evolution of CME flux ropes }


\author*[1,2]{\fnm{Schmieder} \sur{Brigitte}}\email{brigitte.schmieder@obspm.fr}

\author[1,3]{\fnm{Guo} \sur{Jinhan}}\email{jinhan.guo@kuleuven.be}
\equalcont{These authors contributed equally to this work.}

\author[1,4]{\fnm{Poedts} \sur{Stefaan}}\email{stefaan.poedts@kuleuven.be}
\equalcont{These authors contributed equally to this work.}

\affil*[1]{\orgdiv{CmPA/Dept.\ of Mathematics}, \orgname{KU Leuven}, \orgaddress{\street{Celestijnenlaan 200B}, \city{Leuven}, \postcode{3001}, \country{Belgium}}}

\affil[2]{\orgdiv{LIRA}, \orgname{Observatoire de Paris}, \orgaddress{\street{5 place Janssen}, \city{Meudon}, \postcode{92195}, \state{Hauts-de-Seine}, \country{France}}}

\affil[3]{\orgdiv{School of Astronomy and Space Science}, \orgname{Nanjing University}, \orgaddress{\street{163 Xianlin Road, Qixia District}, \city{Nanjing}, \postcode{210023}, \country{China}}}

\affil[4]{\orgdiv{Institute of Physics}, \orgname{University of Maria Curie-Sk{\l}odowska}, \orgaddress{\street{ul.\ Marii Curie-Sk{\l}odowskiej 1}, \postcode{20-031}, \city{Lublin}, \country{Poland}}}


\abstract{Filament eruptions and coronal mass ejections are physical phenomena related to magnetic flux ropes carrying electric current. A magnetic flux rope is a key structure for solar eruptions, and when it carries a  
{ southward magnetic field component when propagating to the Earth. } It is the primary driver of strong geomagnetic storms. As a result, { developing} a numerical model capable of capturing the entire progression of a flux rope, from its inception to its eruptive phase, is crucial for forecasting adverse space weather. The existence of such flux ropes is revealed by the presence of sigmoids in active regions or hot channels by observations from space and ground instruments. After proposing cartoons in 2D, potential, linear, non-linear-force-free-field (NLFFF) and non-force-free-field (NFFF) magnetic extrapolations, 3D numerical magnetohydrodynamic (MHD) simulation models were developed, first in a static configuration and later dynamic data-driven MHD models using high resolution observed vector magnetograms. This paper reviews  { a few} recent developments in data-driven models, such as the time-dependent magneto-frictional (TMF) and thermodynamic magnetohydrodynamic (MHD) models. Hereafter, to demonstrate the capacity of these models to reveal the physics of observations, we present the results for three events { explored in our group}:
1.~the eruptive X1.0 flare on 28 October 2021; 2.~the filament eruption on 18 August 2022; and 3.~the confined X2.2 flare on 6 September 2017. These case studies validate the ability of data-driven models to retrieve observations, including the formation and eruption of flux ropes, 3D magnetic reconnection, CME three-part structures and the failed eruption. Based on these results, we provide some arguments for the formation mechanisms of flux ropes, the physical nature of the CME leading front, and the constraints of failed eruptions.}





\maketitle

\section{Introduction}\label{sec1}

\begin{figure}[ht]
\centering
\includegraphics[width=0.99\textwidth]{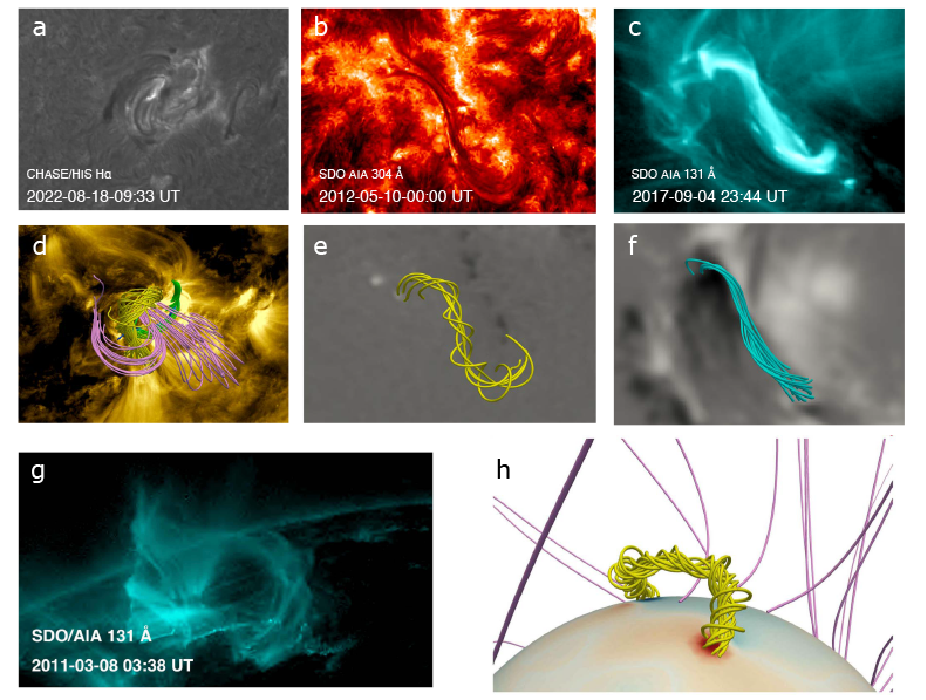}
\caption{Flux rope (FR) observations and corresponding models. Top row: filaments observed on the disk `(a) with CHASE, (b) with SDO/AIA in 304 \AA, (c) sigmoid observed with SDO/AIA in 131 \AA\. Middle row: (d) NLFFF extrapolation of magnetic field (adapted from \citep{Guojhb2023}), (e) inserted flux rope (adapted from \citet{Guo2021}), (f) FR from MHD simulation similar to the observed sigmoid in panel (c) (adapted from \citep{Guojhb2023}). Bottom row: (g) hot channel observed at the limb \citep{Zhang2012, Cheng2013},(h) MHD simulation of a FR (adapted from \citet{Guojh2024b}).
}\label{fig1}
\end{figure}

Solar eruptions release a large amount of magnetic energy into the solar atmosphere and potentially jeopardise the environment of the interplanetary space. Magnetic energy stored in the solar atmosphere is converted into thermal energy, including emissions across the electromagnetic spectrum, kinetic energy with magnetized plasma ejections called coronal mass ejections (CMEs), and ejections of high-energetic particles (SEPs). The origin of CMEs is identified as  { eruptions of filaments}, sigmoids in active regions, and hot channels (Figure~\ref{fig1}) \citep{Green2011, Cheng2011, Schmieder2013, Cheng2016}. 

\begin{figure}[ht]
\centering
\includegraphics[width=0.99\textwidth]{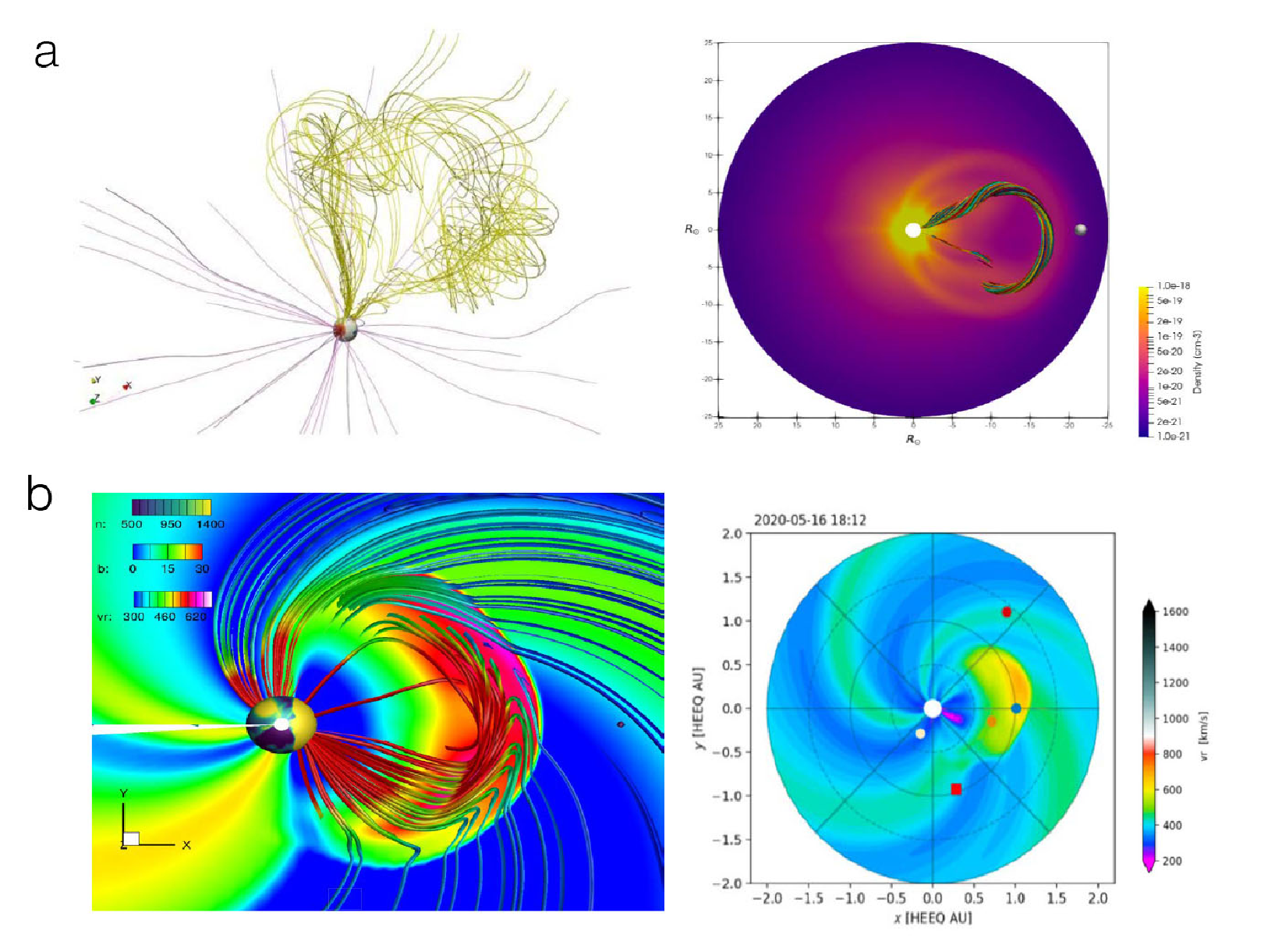}
\caption{Simulations of coronal mass ejections as flux ropes (FR); (a) (left) Flux rope (yellow tubes) in the ambient solar wind (pink tubes) provided by the COCONUT code (adapted from \citep{Guojh2024b}), (right) Density distribution of a FR in the equatorial plane in log scale \citep{Linan2023},  (b) (left) Magnetic cloud EUHFORIA simulation with FR3D (adapted from \citep{Maharana2022}). The charts of colours indicate the values of density (d),  magnetic field (B), and radial velocity (Vr). (right) Equatorial section of a CME in the heliosphere between the Sun and Earth \citep{Maharana2022}.
}\label{fig_CME}
\end{figure}

CMEs are commonly modelled by magnetic flux ropes (FRs) \citep{Zurbuchen2006, Duan2019,Maharana2022}, 
{ defined as a coherent group of magnetic field lines winding an axis with more than one turn or by  }
 a volume channel full of electric currents formed
before or during eruptions (Figure~\ref{fig_CME}). 
Modelled flux ropes  
should be consistent with 
their proxies in observations, such as filaments and hot channels, { which are not by themselves a proof of the existence of the FR.  Sheared arcades can also have dips and support filaments \citep{GuoFR2010}. }Depending on their velocity, CMEs travel through the heliosphere and may produce geomagnetic events after one to 5 days. Therefore, it is essential to forecast them as early as possible \citep{Maharana2022}. Many attempts have been made to find the progenitors of eruptions close to the sun. 
Progenitors of CMEs are detected by kinematics properties such as a slow rise or oscillation of a filament \citep{Zhou2016, Syntelis2016, Joshi2017, Chenpf2008, Ni2021, Cheng2023}, by thermal signatures with very hot channels \citep{ChengFR2011, Zhang2012} and by magnetic flux, e.g., twist 
increase and magnetic helicity fluctuations { \citep{Webb2000, Guo2013, Pariat2017, Moraitis2019,Kusano2020}.}

\begin{figure}[ht]
\centering
\includegraphics[width=0.99\textwidth]{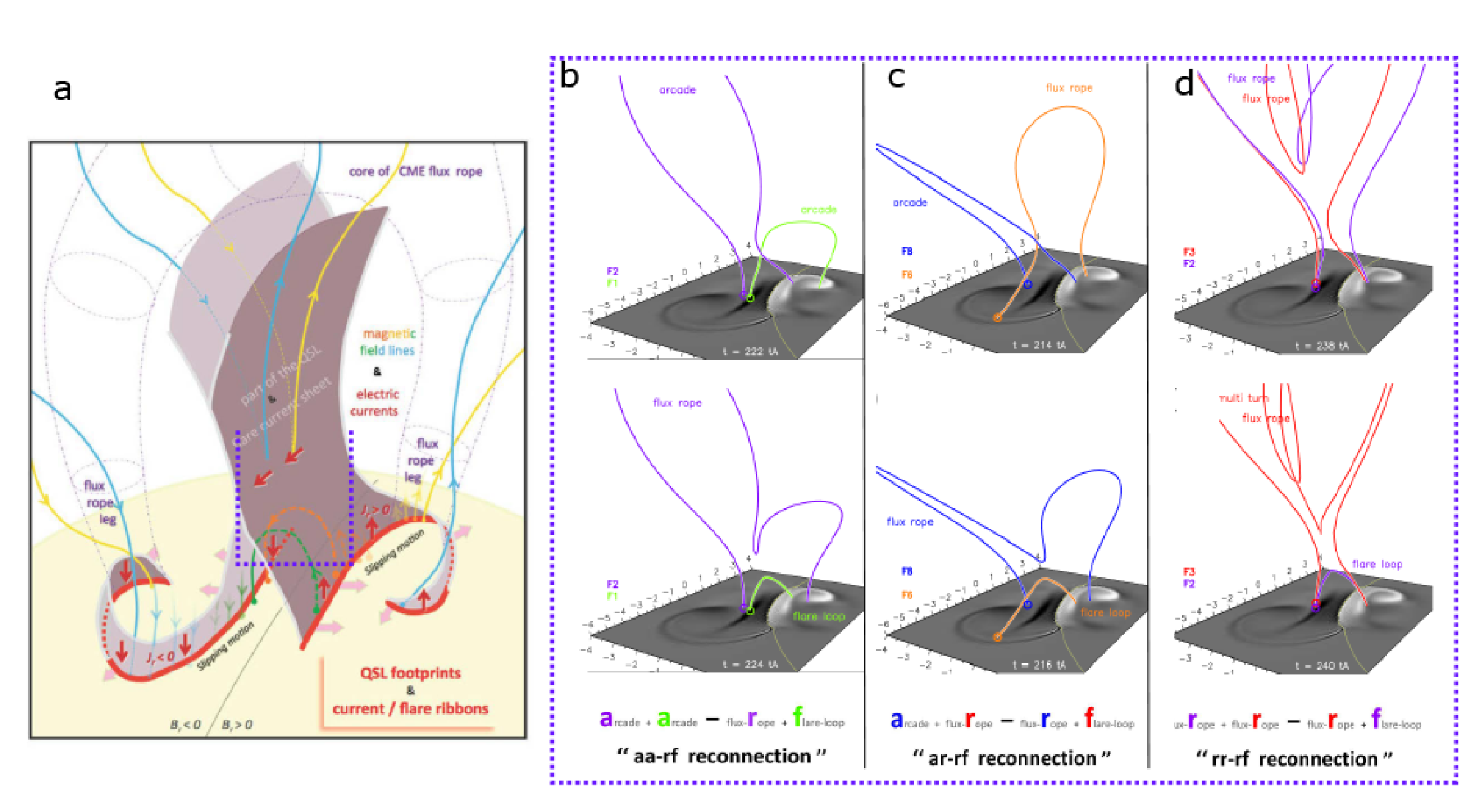}
\caption{(left panel) Sketch of the  3D magnetic structures and magnetic field lines found in eruptive active regions modelled with the OHM simulation \citep{Aulanier2005, Janvier2014}; (right panels), the different possibilities of a flux rope to reconnect with arcades, and overlying magnetic field \citep{Aulanier2019}.
}\label{cartoon}
\end{figure}

Looking at sunspots and observing scarves in the penumbra and umbra could help to detect the ``feet'' of the flux rope in an eruption stage \citep{Xing2024}. Figure~\ref{cartoon} ~(panel a) represents a sketch of the 3D magnetic structures and magnetic field lines found in eruptive active regions modelled with the OHM simulation \citep{Aulanier2005, Janvier2014};  the hooks of the ribbons  are  interpreted as the footprints of the flux rope of the eruption.
Another proxy is the observation of hot channels in the corona using the AIA imager on board the Solar Dynamics Observatory (SDO) in hot temperature filters (131\;\AA\  and 94\;\AA) \citep{Cheng2023}. It was shown that the morphology of the flux rope with a small or large radius observed on the disk is also a precursor for small or large storms 
\citep{Guojh2024b}.

Solar eruptions have been explained in a first attempt with the 2D standard model (CSHKP) developed between 1964 and 1976 by \citet{Carmichael1964}, \citet{Sturrock1966}, \citet{Hirayama1974}, \citet{Kopp1976}, based on magnetic reconnection in the legs of flux rope or arcades evolving { in 
magnetohydrodynamic (MHD) environment }\citep{Lin2000}. Magnetic reconnection occurs in non-ideal, highly conducting plasmas where the magnetic field lines are generally frozen to the plasma. The key processes are well present in these standard models, but still, many observable features in 3D are not taken into account. Some magnetic properties cannot be described in the 2D standard model, such as the toroidal and poloidal flux and the twist of the magnetic field in the flux rope. In 2D models, magnetic reconnection occurs in the current sheet induced by a null point. However, the null point becomes unnecessary in the 3D scenario, and magnetic reconnection occurs in the region where magnetic field connectivity changes drastically, namely, the quasi-separatrix layers (QSLs) \citep{Priest1995}. The reconnection in QSLs is generally called slipping reconnection, in which the slipping velocity of one field line increases with the local norm $N$ \citep{Janvier2013}. { QSLs are regions of high distortion of the mapping of magnetic field lines anchored in the photosphere \citep{Priest1995,Demoulin1996}.  This distortion is described by derivatives of the field line mapping functions expressed via Jacobian matrices of the MHD equations, and the vertical norm $N$  is directly related to the squashing factor.} 

QSLs are { layers where current flows easily in response to changes in the plasma.  They  may contain } high electric current-density 
if neighbouring magnetic field lines can 
change their different footprint locations drastically. The squashing degree $Q$ is a parameter that indicates the gradient of connectivity change in the magnetic field volume under consideration \citep{Titov2002}. A high squashing factor $Q$ suggests that the magnetic field is more distorted and that it is more likely to { form} high electric current densities. { Same as the X point in 2D, the 3D reconnection most likely starts in  Hyperbolic Flux Tubes (HFT) involving two intersecting QSLs,}  which comprise two intersecting layers \citep{Titov2002, Zhao2014}. Each layer arises from a crescent-shaped strip with one pole and tapers toward the other. The crescent-shaped bands connecting two sunspots have the same polarity. The intersections of these domains with the photosphere are characterised by emission enhancements in all wavelength ranges, such as flare ribbons visible in UV and optical ranges and kernels in X-rays during flares and eruptions. The cartoon in Figure~\ref{cartoon} (top left panel) summarises these magnetic structures observed in 3D and retrieved in the  {  Observationally driven High-order Magnetohydrodynamics  (OHM) simulations  with
the line-tying and zero-$\beta$ approximations  \citep{Aulanier2005,Aulanier2010,Janvier2014, Aulanier2019}).}

Several 3D MHD simulation models with zero-$\beta$ assumption have been developed nowadays both in bipolar magnetic configuration \citep{Amari2003, Kliem2006, Aulanier2005, Aulanier2010} and in multi-polar configurations \citep{Antiochos1999}. { In the bipolar model, 
flux cancellation or shearing motions can build up a flux rope, which is subject to the kink instability or the torus instability leading to the eruption of the FR}  \citep{Schmieder1996, Kliem2006, Jiang2021, Guo2019}.  In the multi-polar configuration, \citet{Antiochos1999} and \citet{Chen2000} proposed the breakout model and emerging flux trigger, respectively.

These models are compelling and recover the main features of flares, incl.\ the flare ribbons, the post-flare loops, and the site of reconnection \citep{Janvier2013}. Moreover, in  the OHM simulations 
\citep{Aulanier2019}, slipping magnetic reconnection occurs in the HFT structure below the eruptive flux rope. 
In their simulations, three types of reconnection geometries are recognized: { the reconnection $(aa-rf)$ refers to reconnection between an Arcade and another Arcade leading to the formation of a flux Rope and a Flare loop, $(rr-rf)$ reconnection between a flux Rope and another flux Rope leading to a flux Rope and a Flare loop, {\bf  $(ar-rf)$ } reconnection between an Arcade and a flux Rope leading to a flux Rope and a Flare loop.}
These reconnection geometries can explain many complicated 3D observational phenomena, such as the shift of filament legs \citep{Dudik2019}, saddle-like flare loops \citep{Juraj2021}, footpoint drifting and decrease in toroidal fluxes of CME flux ropes \citep{Xing2020}. Additionally, this theoretical model can be applied to ample parameter space, including stars, e.g.\ to forecast super flares \citep{Aulanier2013}.

However, in the zero-$\beta$ models \citep{{Aulanier2010,Kliem2013,Inoue2018,Natur.554..211A,Aulanier2019,Zhong2021}}, 
or in isothermal MHD models \citep{Jiang2016, Jiang2018},
 the thermal properties of the plasma are discarded or at least drastically simplified. Therefore, comparing these simulation results with the multi-temperature images of SDO/AIA is difficult.

Recent studies are focused directly on the observations, which leads to data-driven or data-constrained models { (see the reviews of \citet{Inoue2018,Jiang2022})}. In data-driven models, the observational data from the photosphere are taken as inputs for driving the coronal field and the related plasma flows, such as the magnetic fields \citep{Jiang2016}, velocity \citep{Hayashi2018, Jiang2021, Kaneko2021}, the combination of the velocity and the magnetic field \citep{Guo2019, Guojha2023}, and the electric field \citep{Cheung2012, Hayashi2018, Pomoell2019, Fisher2020, Afanasyev2023}. The outcomes of these data-driven or data-constrained models are directly comparable to multi-wavelength observations, demonstrating significant potential in quantitatively elucidating the fundamental physical mechanisms behind the observations \citep{Jiang2016,Jiang2022,Guojha2023}.

This review is interested in data-constrained and data-driven models exploiting velocity and magnetic field data.
Section~2 details the data used to drive the models. Section~3  demonstrates how data-constrained models are achieved in a few case studies.
Section~4 is focused on data-driven models and the formation of flux ropes. Section~5 explains the role of the magnetic tension leading to confined flares.

\section{Data inputs for MHD models}

\subsection{Extrapolation of the magnetic field}
Numerous recent studies involving numerical models based directly on observational data have been conducted.
The critical question is the distributions of the magnetic field and the electric current in the corona.

Solar magnetographs have been developed to produce photospheric magnetograms from the ground and space with increasingly higher spatial and temporal resolution. { The full disk of the magnetograph  MDI on board SOHO since 1996 \citep{Scherrer1995}  was already a progress, even with only the line-of-sight magnetograms,  since 2011  HMI on board SDO  \citep{Scherrer2012} allows us to have magnetic field vector-maps 
every 12 min.} The technique of inversion of the Stokes parameters is well developed for HMI data using the Very Fast Inversion of Stokes Vector or UNNOFIT \citep{Borrero2011, Bommier2016}, which are Milne–Eddington based algorithms. A minimum energy method \citep{Metcalf1994, Leka2009, Leka2022}  is used to resolve the $180^\circ$ ambiguity in the transverse field \citep{Metcalf1994, Leka2009}.  
The SDO/HMI vector magnetograms must be pre-processed to ensure that the photospheric magnetic field satisfies the NLFFF model assumptions in the local Cartesian coordinate system, as the photosphere is not always force-free. The pre-processing follows the methods developed by \citet{Wiegelmann2006} and discussed by \citet{Valori2010} and \citet{Thalmann2019}. All these steps are mathematically not well posed, therefore extrapolations is a difficult task. Measurement of physical parameters can show an ambiguity in the results \citep{Thalmann2019}.  Moreover, such pre-processing does not include the projection effects, and central disk eruptions are often considered. Since recently, pre-processing involves correcting the projection effects and removing the Lorentz force and torque \citep{Guo2017}. 
  
Different methods to improve the magnetic field extrapolation in the corona have been developed using potential field \citep{Chiu1977}, linear-force-free (LFF) \citep{Aulanier1998, Mandrini2014}, non-linear-force-free-field (NLFFF) \citep{Guo_jet2013, Wiegelmann2021} assumptions.  In these studies, the models are restricted to static reconstructions of the nearly force-free coronal magnetic field. 
With only one magnetogram, obtained just before the eruption, two sets of magnetic field lines are drawn, one corresponding to the magnetic field before the eruption and one corresponding to the magnetic field after reconnection \citep{schmieder1997}. The large-scale magnetic field configuration does not change significantly during typical solar and confined flares. The observed flare loops involved both before and after the flare can thus be fitted using a single magnetogram \citep{Mandrini1996, Dalmasse2015, Green2017, Zuccarello2015, Joshi2019, Guojha2023, Guojhb2023}.
Notice that just before the eruption, the field is already well-developed and, by definition, unstable. Hence, using magnetograms immediately preceding the eruption for the coronal magnetic field reconstruction and taking this magnetic field as the initial MHD simulation condition reproduces the erupting field's fast dynamic phase. { The evolution of coronal magnetic fields in the pre-eruption phase and the triggering of the eruption are, however, not revealed in such simulations. It is well accepted that ideal MHD instability and magnetic reconnection are responsible for initiating a CME. Regarding the ideal MHD instability, it incorporates kink instability \citep{Hood&Priest1981} and torus instability \citep{Kliem2006}, wherein the former are controlled by the twist number of the pre-eruptive flux rope, while the latter is determined by the decaying degree of the overlying background magnetic fields. As for the role of magnetic reconnection in leading to a CME, it is classified into tether-cutting \citep{Moore1980, Jiang2021}, emerging flux \citep{Chen2000} and breakout \citep{Antiochos1999} models. \citet{Jiang2018} and \citet{Duan2019} investigated the role of ideal MHD instability in triggering solar eruptions by computed the twist number and decay index of the pre-eruptive magnetic fields. They found that the threshold of these two metrics can be adopted to distinguish confined and eruptive flares to a great extent, and some exceptions could be due to magnetic reconnection. In particular, \citet{Jiang2024} detailed the fundamental role of magnetic reconnection in triggering solar eruptions.}
Thus, such models do not allow for the identification of the actual trigger and dynamic evolution of solar eruptions. 

\subsection{MHD relaxation Model}

In the previously mentioned studies, one magnetogram or a series of magnetograms was used, but the eruption mechanism could only be investigated tentatively because no dynamics were included. Even a time sequence of magnetic fields reconstructed following the coronal evolution does not reflect its intrinsic dynamics because these magnetic fields are treated as independent. By definition, the reconstructed coronal magnetic field immediately before the eruption is unstable. 

It is necessary to relax the solar atmosphere, an electrically conductive fluid, to some minimum energy state to study flares \citep{Yeates2020}.  
Potential field extrapolations are often used. These are minimum-energy models for the coronal magnetic field of the Sun. NLFFF extrapolations are therefore performed using an MHD relaxation method. Different methods exist we may quote the \citet{Zhu2013,Zhou2016}, \citet{Jiang2013} and \citet{Guo2016a, Guo2016b} methods. The method described by \citet{Zhu2013} and \citet{Zhou2016} consists of computing the magnetohydrostatic state of the solar atmosphere. This relaxation is achieved in several case-study events \citep{Joshi2019}. { 
The CESE–MHD–NLFFF model developed by \citet{Jiang2010} is based on an MHD-relaxation method
which seeks approximately force-free equilibrium. It solves a set
of modified zero-$\beta$ MHD equations with a friction force using an
advanced conservation-element/solution-element (CESE) space-
time scheme on a nonuniform grid with parallel computing. With this method, 45 flares have been analysed, showing that flux ropes exist in the pre-flare phase, and by computing the twist parameter and the decay index, they show that eruptive and confined flares can be forecast \citep{Duan2019}}.
The Guo method is based on the magneto-friction (MF) relaxation method (see next subsection).

\subsection{Magneto-frictional method}
The magneto-friction method is a simplification of the MHD model, which omits gravity and thermal pressure, and the velocity is assumed to be proportional to the local Lorentz force. As such, the MF method only computes the magneto-induced equation, and the final relaxed state will be converted to a force-free field. The governing equations of the MF relaxation are as follows:

\begin{eqnarray}
 && \frac{\partial \boldsymbol{B}}{\partial t} + \nabla \cdot(\boldsymbol{vB-Bv})= -\nabla \times(\eta \boldsymbol{j}),\label{eq——MF1}\\
 && \boldsymbol{v}=\frac{1}{\nu}\ \frac{\boldsymbol{j \times B}}{ B^{2}},\label{eq-MF2}\\
 \notag  
 \end{eqnarray}
where $\nu$ is the viscous coefficient of the friction and $\eta$ is the magnetic diffusivity. Hence, this method can extrapolate static coronal NLFFF from the potential field model where the bottom boundary is provided by the observed vector magnetograms in the photosphere, such as \citet{Guo2016a}. Additionally, to reduce computing expenses and simultaneously obtain the temporary evolution of 3D coronal magnetic fields, many authors have adopted a time-dependent magnetofrictional (TMF) model to perform data-driven models \citep{Cheung2012, Cheung2015, Pomoell2019, Kilpua2021, Lumme2022,  Afanasyev2023, Guojh2024}, where the bottom boundaries are provided by a series of observed magnetograms or their derived electric fields. Compared to other methods for NLFFF extrapolation, such as the optimisation \citep{Wheatland2000, Wiegelmann2004} and Grad-Rubin \citep{Sakurai1981, Amari2006, 1977ApJ...212..873C} methods, the numerical computation of the magneto-frictional relaxation is still based on the numerical schemes of the MHD equations. As a result, it is easier to perform in open-source codes for MHD numerical simulations and coupled with some advanced numerical strategies, i.e., the adaptive mesh refinement (AMR) and stretching grids, constrained-transport (CT) method for { keeping 
magnetic-field divergence freeness} introduced in the computation process, and magnetic-field splitting to decrease the numerical diffusion.


\citet{Guo2016b} made a significant step forward in this domain by the implementation of a magneto-frictional module in the Message Passing Interface Adaptive Mesh Refinement Versatile Advection Code \citep[MPI-AMRVAC\footnote{http://amrvac.org},][]{Xia2018, Keppens2023}. The magneto-frictional method has also been applied to several case studies. Thus, its applicability has been demonstrated in Cartesian as well as in spherical coordinates and both uniform and block-adaptive octree grids \citep{Guo2016a, Zhong2019, Guo2021}.
Moreover, in NLFFF modelling, local high spatial resolution can be achieved simultaneously with a large field-of-view using parallel and block-adaptive magneto-frictional relaxations \citep{Guo2019, Guojha2023, Guojhb2023}.
For example, in the paper by \citet{Guojhb2023} concerning the non-radial motion of a filament during its eruption, observed 
by the Chinese H$\alpha$ Solar Explorer (CHASE/HIS; \citep{Li2019, Li2022}),
the initial magnetic field was provided by a NLFFF model with a multi-step construction procedure. In the first step, an SDO/HMI vector magnetogram was pre-processed to ensure that the photospheric magnetic field agrees with the NLFFF model assumptions in the local Cartesian coordinate system. In the second step, the 
MF in MPI-AMRVAC was applied and succeeded in producing an excellent initial condition for the MHD simulation of the eruption 
(Figures~\ref{fig1} panel a and d).

\subsection{Magnetic flux ropes}

With the NLFFF extrapolations, it is not apparent that bundles of highly twisted magnetic field lines can be obtained that prove the existence of a flux rope. Magnetic field lines may not show a flux rope but rather an arcade \citep{GuoFR2010}. In these cases, flux ropes should be formed during the eruption process so they can be produced during the relaxation process. 

For example, in the study of \citet{Prasad2023}, the MHD simulation starts with an extrapolated non-force-free magnetic field that is generated from a photospheric vector magnetogram of the concerned active region taken a few minutes before the onset of the flare. A sheared arcade is observed along the polarity inversion line in the magnetic topology before the flare. The shear created by the footpoints anchored in the photosphere initiates tether-cutting magnetic reconnection, which subsequently produces a flux rope above the flare arcade.
The rising flux rope forms in a torus-unstable region, explaining the eruption. Similarly, \citet{WangX2023} reproduced the formation of a pre-eruptive magnetic flux rope in NOAA 11429 and its eruption due to torus instability by implementing the photospheric velocity field.
In another study, presented by \citet{Jiang2016}, after the analysis of the magnetic field topology, the transition from the pre-eruptive to the eruptive state is probably due to the upward expansion of internally stressed magnetic arcades of newly emerged flux reconnecting with external magnetic field which is the trigger of the eruption. The other possibility is that the coronal magnetic flux rope cannot be constructed well if the information is not correctly transformed from the bottom boundary to the coronal in weak fields. The NLFFF extrapolation is, in fact, an ill-posed problem \citep{Low1990}. To this end, the idea of incorporating the information from the coronal observations for reconstructing coronal magnetic fields is proposed. For example, a possibility is to insert a flux rope mimicking the shape of the eruptive filament as has been suggested already 20 years ago by \citet{Ballegooijen2004}, and as was more recently achieved \citep{Bobra2008, Su2009, Su2015, Mackay2020, Guojha2023, Guojhb2023}. In those cases, the relaxation can be done after the insertion of the flux tube. 

In the study of \citet{Guojhb2023} concerning the eruption of an H$\alpha$ filament observed by CHASE/HIS 
and the Atmosphere Imaging Assembly  \citep[AIA,][]{Lemen2012}  on board the  Solar Dynamics Observatory (SDO), the potential field is first extrapolated using Green's function with the $B_z$ component \citep{Chiu1977} after the pre-processing of a magnetogram. Then, the flux rope is superposed onto it. This flux rope is constructed with the Regularised Biot-Savart laws (RBSLs; \citet{Titov2018}). The RBSL method proposed by \citet{Titov2018} can construct flux ropes with the axis of arbitrary path, which are more consistent with complicated flux rope proxies in observations. Based on this method, \citet{Guojhb2023} construct the flux rope structure for a filament observed by CHASE (Figures~\ref{fig1}a and \ref{fig1}d), in which the axis path, toroidal flux and cross-section radius of the RBSL flux rope are measured from observations. In their pipeline, the parameters of the RBSL flux rope are fully derived from or based on the observations. The flux rope path and its minor radius are approximated by the filament path and width \citep{Guojh2022}, respectively, and the toroidal flux is approximated using the $B_z$ map (refer to \citet{Guo2019} for more details). 
They calculated the RBSL flux rope two times. The first time, they only computed the bottom boundary to prepare the boundary condition of the potential field extrapolation to keep the consistency of the superposed magnetic fields with the observed magnetogram. The second time is to compute the 3D distribution of the RBSL flux-rope magnetic fields in constructing the NLFFF.

\begin{figure}[ht]
\centering
\includegraphics[width=0.99\textwidth]{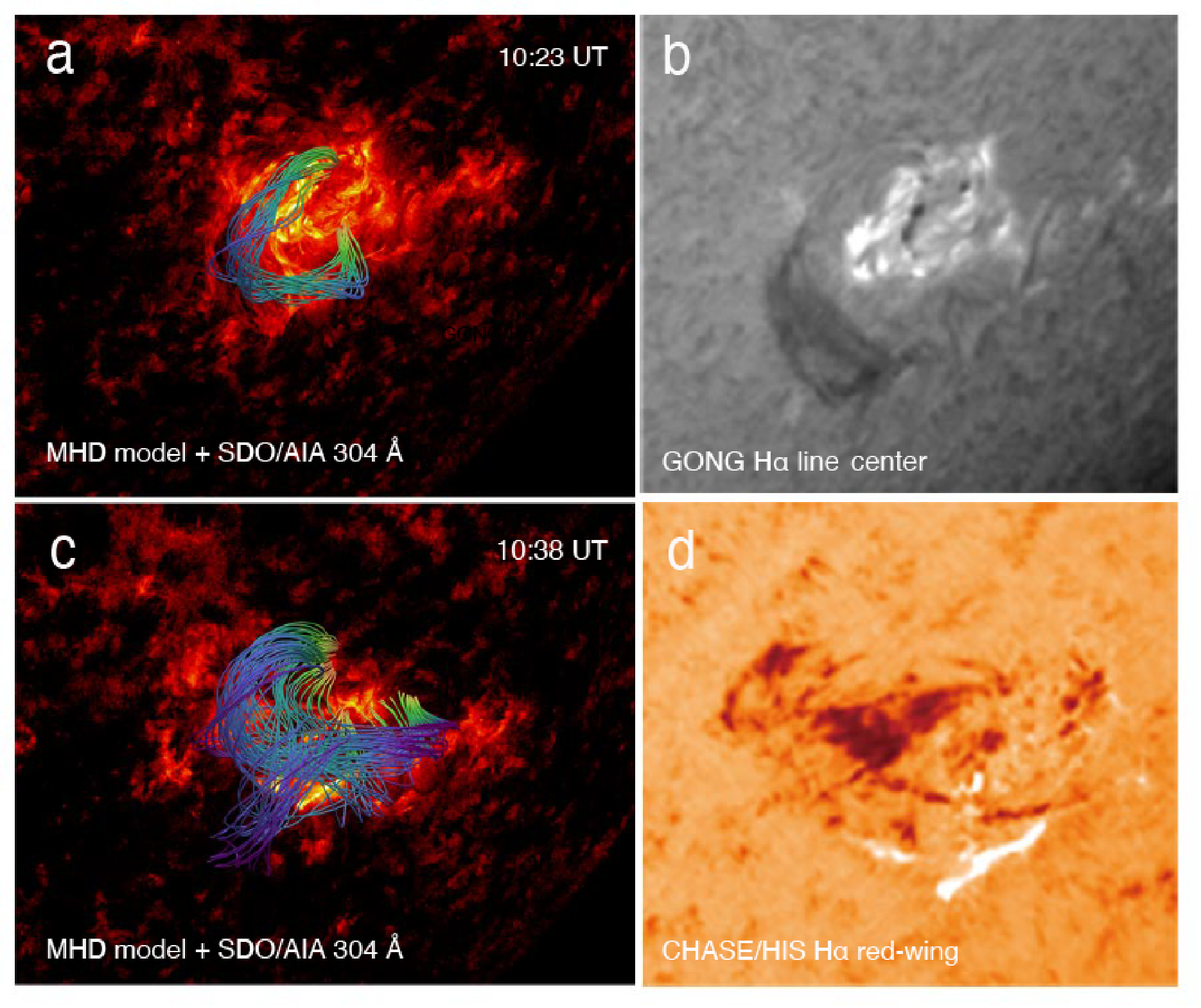}
\caption{Comparison between the modelled flux rope and the filament observed in SDO/AIA 304~\AA\  (Panels a and c), GONG H$\rm \alpha$ line-centre (Panel b) and CHASE/HIS H$\rm \alpha$ red-wing image (Panel d) (adapted from  \citet{Guojhb2023}).
}\label{chase}
\end{figure}

To not perturb the photospheric magnetic field by the insertion of the flux rope, \citet{Guo2019, Guoy2021} subtracted the photospheric magnetic field of the RBSL flux rope from the observed $B_z$ before extrapolating the potential field. The resulting superposed photospheric magnetic fields combine the potential and RBSL flux-rope magnetic fields and agree with the observed surface magnetic field. Finally, the magneto-frictional method relaxes this magnetic field to a force-free state. It is what has been done in recent studies \citep{Guo2019, Guojh2021, Guojha2023, Guojhb2023}. The flux rope evolving with time obtained by the subsequent MHD simulation can be compared with the evolution of the observed filament (Figure~\ref{chase}).

\begin{figure}[ht]
\centering
\includegraphics[width=0.99\textwidth]{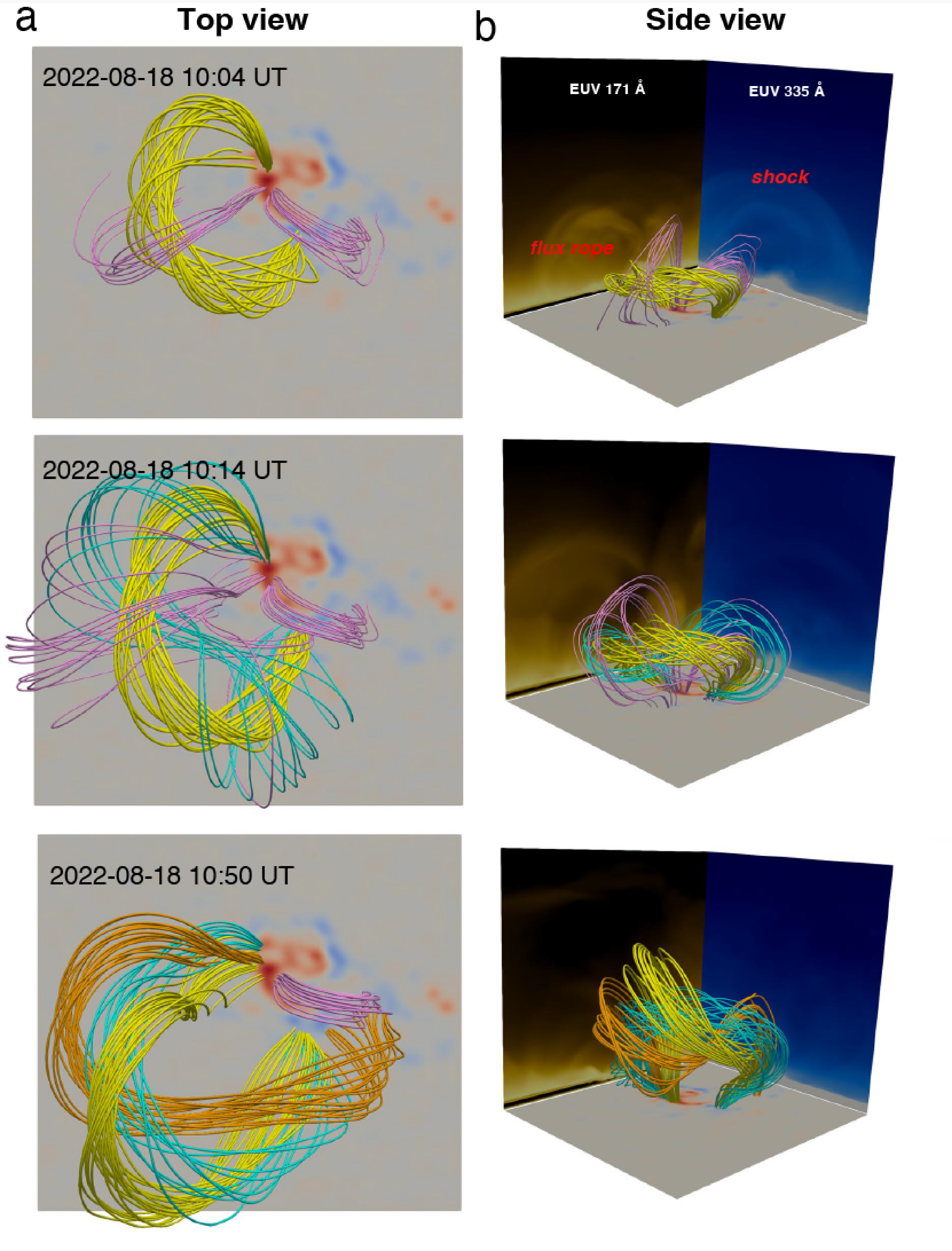}
\caption{Top (left panels) and side (right panels) views of eruptive flux ropes obtained by a data-constrained MHD simulation. The synthesised EUV 171~\AA\ and 335~\AA\ images are shown in side views for comparison with SDO/AIA observations (adapted from \citet{Guojhb2023}).
}\label{chase_TMF}
\end{figure}

\begin{figure}[ht]
\centering
\includegraphics[width=0.99\textwidth]{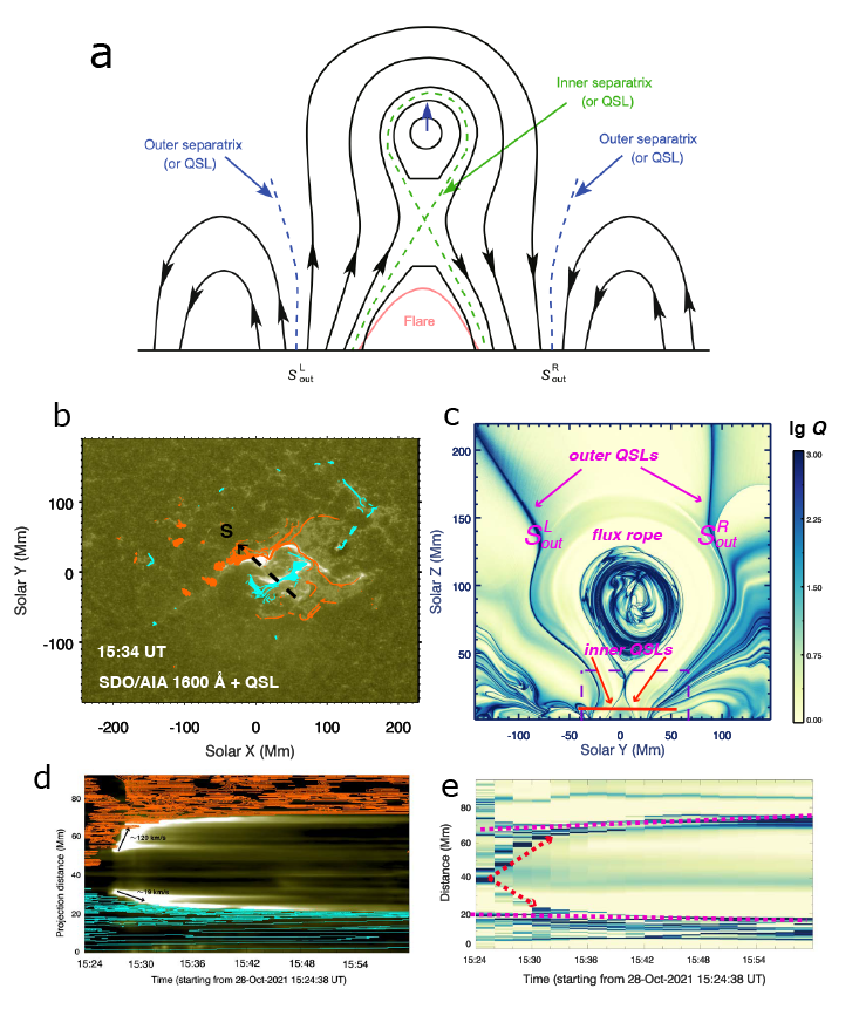}
\caption{The stopping position of flare ribbons revealed by data-driven MHD simulation. Panel (a) shows the sketch of magnetic reconnection considering inner and outer QSLs, in which the separating inner QSLs correspond to flare ribbons in observations, and the outer QSLs determine the stopping position of flare ribbons. Panels (b)--(d) show the results of data-driven MHD simulation, validating the effects of inner and outer QSLs in explaining the dynamic evolution of flare ribbons. Panel (b) illustrates the $Q$ distribution in the photosphere, overlaid on SDO/AIA 1600~\AA\ observations. The black dashed line in panel (b) shows the slice for the time-distance diagram in panel (d). Panels (c) and (e) show the $Q$ distribution on the side plane and dynamic evolution, respectively (adapted from  \citet{Guojha2023}).
}\label{ribbon}
\end{figure}
\section{Thermodynamic MHD models}
The zero-$\beta$ models and even the isothermal models cannot self-consistently synthesise the radiation from the density and temperature, and the results from these models are thus not directly comparable to EUV observations. Hence, data-driven thermodynamic MHD models need to be developed to a better insight into the nature of some emission structures like filaments, EUV waves, and coronal loops \citep{Guojha2023}.
Recently, a solar flare and a CME have been reproduced by a data-based MHD model considering the non-adiabatic effects \citep{Fan2022}.  

There are two types of models: data-driven and data-constrained. The initial magnetic field is usually reconstructed using the NLFFF or magnetostatic extrapolation method, and the photospheric or low-coronal boundary is fixed or provided by numerical extrapolations. In these cases, a pre-eruptive magnetic flux rope generally exists to generate solar eruptions, as shown in \citet{Guoy2021}. In real data-driven models, 
the long-term evolution of the active region is studied to model the flux rope's formation (see Section~4).

\subsection{MHD and thermodynamic equations}\label{sec:met}

In two recent works, \citet{Guojha2023} and \citet{Guojhb2023} perform thermodynamic simulations of the eruptive events on 28 October 2021 and 18 August 2022, respectively. In \citet{Guojha2023}, they adopted a nonadiabatic MHD model that considers the field-aligned thermal conduction, empirical heating, and the optically thin radiation losses in the corona. The governing equations read as follows:

\begin{eqnarray}
 && \frac{\partial \rho}{\partial t} +\nabla \cdot(\rho \boldsymbol{v})=0,\label{eq1}\\
 && \frac{\partial (\rho \boldsymbol{v})}{\partial t}+\nabla \cdot(\rho \boldsymbol{vv}+p_{_{tot}}\boldsymbol{I}-\frac{\ \boldsymbol{BB}}{\mu_{0}})=\rho \boldsymbol{g},\label{eq2}\\
 && \frac{\partial \boldsymbol{B}}{\partial t} + \nabla \cdot(\boldsymbol{vB-Bv})=0,\label{eq3}\\
&& \frac{\partial \varepsilon}{\partial t}+\nabla \cdot(\varepsilon \boldsymbol{v}+p_{_{tot}}\boldsymbol{v}-\frac{\boldsymbol{BB}}{\mu_{0}}\cdot \boldsymbol{v}) =\rho \boldsymbol{g \cdot v}+H_{0}e^{-z/\lambda}-n_{\rm e}n_{\rm _{H}}\Lambda(T) \notag \\             && + \nabla \cdot(\boldsymbol{\kappa} \cdot \nabla T),\label{eq4}
\end{eqnarray}
where $p_{_{tot}} \equiv p + B^2 / (2\mu_{0})$, corresponds to the sum of the thermal pressure and the magnetic pressure, $\boldsymbol{g}=-g_{\odot}r_{\odot}^2/(r_{\odot}+z)^2\boldsymbol{e_{z}}$ denotes the gravitational acceleration, $g_{\odot}= \rm 274\;m\,s^{-2}$ corresponds to the gravitational acceleration at the solar surface, $r_{\odot}$ is the solar radius, $\varepsilon =\rho v^2/2+p/(\gamma -1)+ B^2 / (2\mu_{0})$ is the total energy density, the term $\nabla \cdot(\boldsymbol{\kappa} \cdot \nabla T)=\nabla \cdot(\kappa_{\parallel}\boldsymbol{\hat{b}\hat{b}} \cdot \nabla T)$ represents field-aligned thermal conduction, $\kappa_{\parallel} =10^{-6}\ T^{\frac{5}{2}}\ \rm erg\ cm^{-1}\ s^{-1}\ K^{-1}$ is the Spitzer heat conductivity, $n_{\rm e}n_{\rm _{H}}\Lambda(T)$ is the optically-thin radiative losses, $H_{0}e^{-z/\lambda}$ is an empirical heating to maintain the high temperature of the corona.

In these two papers, \citet{Guojha2023, Guojhb2023} showed that the twisted flux rope and sheared field lines compare well with the observed filament and chromosphere fibrils. This indicates that the NLFFF model can serve as an initial condition for the MHD simulation, as shown in Figures~\ref{fig1}a and \ref{fig1}d. The initial density and pressure then define a hydrostatic atmosphere from the chromosphere to the corona, which is described as follows:
\begin{eqnarray}
 T(z)=  \begin{cases} T_{_{\rm ch}}+\frac{1}{2}(T_{_{\rm co}}-T_{_{\rm ch}})({\rm tanh}(\frac{z-h_{_{\rm tr}}- 0.27}{w_{\rm tr}})+1)\qquad&  z \leq h_{_{tr}},\\ (\frac{7}{2}\frac{F_{\rm c}}{\kappa}(z-h_{\rm tr})+T_{_{\rm tr}}^{7/2})^{2/7} & z > h_{_{tr}}, \end{cases} \label{eq6}
 \end{eqnarray}
where $T_{_{\rm ch}}=8000\;$K corresponds to the chromospheric temperature, $T_{_{\rm co}}=1.5\;$MK represents the coronal temperature, $h_{_{\rm tr}}=2\ $Mm and  $w_{_{\rm tr}}=0.2\;$Mm control the height and thickness of initial transition region, and $F_{\rm c}=2 \ \times 10^{5}\ \rm erg \ cm^{-2}\ s^{-1}$ is the constant thermal conduction flux. Hereafter, the density distribution is calculated from the number density at the bottom, i.e., $1.15 \times 10^{15}\;$cm$^{-3}$.

\subsection{Case-studies of 28 October 2021 and 18 August 2022}
For these two case studies, \citet{Guojha2023, Guojhb2023} compared the results of the simulations to AIA observations. By including thermal conduction and radiative losses in the energy equation, they developed a novel data-driven thermodynamic magnetohydrodynamic model that can capture the thermodynamic evolution in contrast to the previous zero-$\beta$ model. Their numerical model reproduces multiple notable observational eruption features, incl.\ the erupted filament morphology, its path, and the flare ribbons. In the case of the event on 18 August 2022, the simulations indicate that magnetic reconnection of the flux-rope leg with the neighbouring sheared arcades may be the primary mechanism for the lateral drifting of filament materials. This also causes the flux-rope rotation \citep{Guojhb2023}. This conduct agrees with the 3D $ar-rf$ reconnection model \citep{Aulanier2019}. They pointed out that the lateral drifting of filament materials can also serve as an observational signature for $ar-rf$ reconnection, in addition to saddle-like flare loops \citep{Juraj2021} and the shift of filament legs \citep{Dudik2019}. They obtained synthesised images similar to the SDO/AIA observations (Figure~\ref{chase_TMF}).

\citet{Guojha2023} studied the 28 October 2021 event and reproduced the main observational characteristics of the X1.0 flare 
in NOAA active region 12887, starting with the morphology of the eruption and including the kinematics of the flare ribbons, the EUV emission, the CME \citep{Devi2022}, and even the two components of the EUV waves predicted by the magnetic stretching model of \citet{Chenpf2002}, namely a fast-mode shock wave and a slower apparent wave which is due to consecutive magnetic field line stretching. 

\begin{figure}[ht]
\centering
\includegraphics[width=0.99\textwidth]{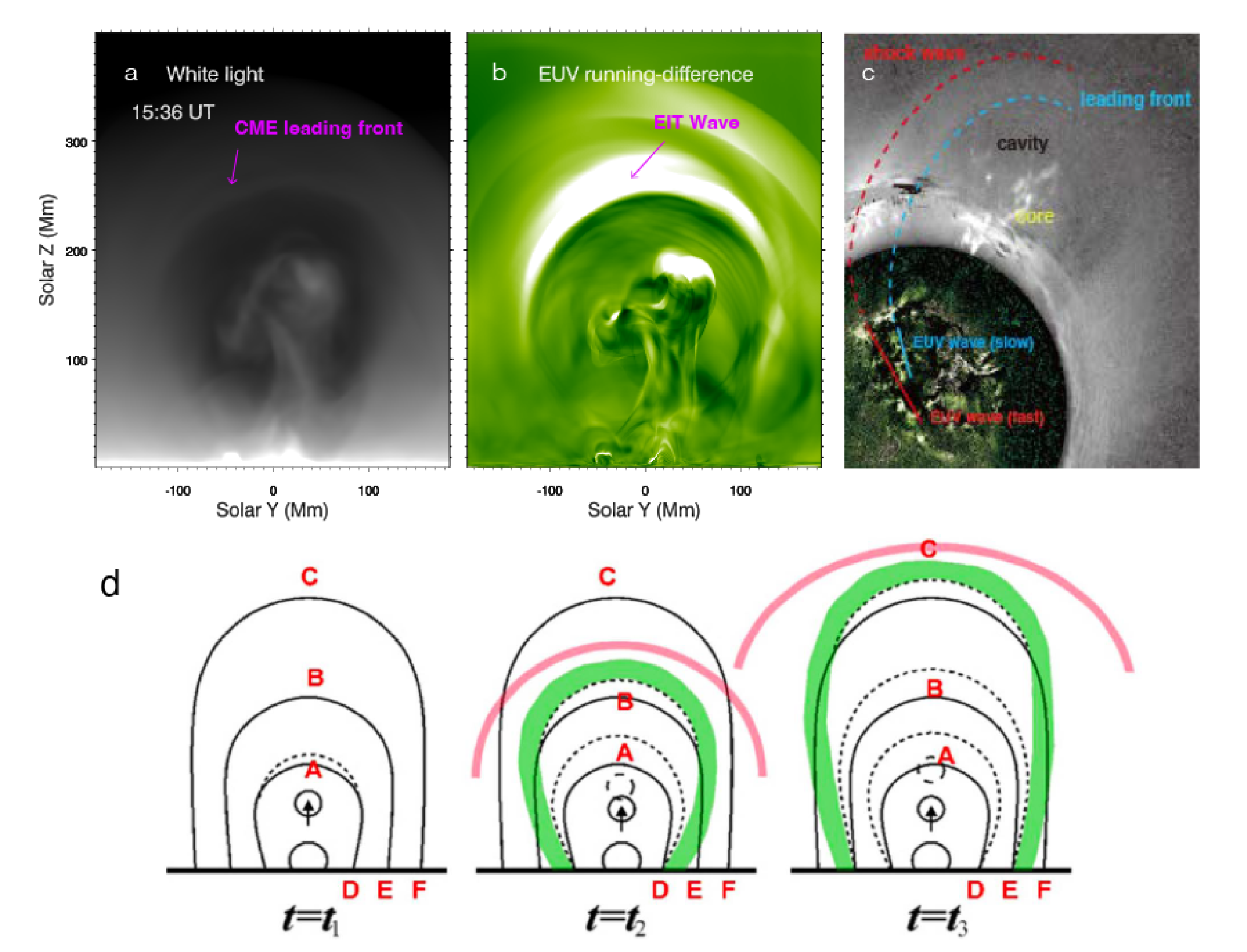}
\caption{CME and EUV wave: (top panels)  comparison between synthesised images (a, b)  obtained with an MHD simulation and (c) observations by STEREO  coronagraph (COR1) and EUV imagery (adapted from \citet{Guojha2023}). The fast- and slow-component of the EUV waves are roughly co-spatial with the shock and the CME frontal edge, respectively, while the EUV dimmings are co-spatial with the CME cavity. Panel (d) illustrated the sketch of the field-line stretching model in explaining the two components of EUV waves and the CME leading front (adapted from \citet{Chen2009}). The results of the data-driven model align with the field-line stretching model prediction. 
}\label{CME}
\end{figure}

The simulation also reveals some fascinating phenomena. The flare ribbons initially separate and eventually stop at the outer stationary QSLs. These QSLs correspond to the borders of the filament channel and demarcate the flare ribbons' final positions. These, in turn, can be used to predict the lifetime and size of a flare before it occurs (Figure~\ref{ribbon}). Moreover, the side views of the synthesised EUV and white-light images display the typical three-part structure of CMEs. The bright leading front is approximately co-spatial with the EUV wave's non-wave component, which is in agreement  with the previous observations \citep{Chen2009}. These simulation results reinforce the effects of the magnetic field-line stretching model in explaining the slow component of EUV waves (Figure~\ref{CME}).


\section{Long-term data-driven Models}

Data-driven models are superior to data-constrained types.
The corona responds to the photosphere in real time.
Data-driven models are suitable for studying the long-term evolution of active regions.

The NOAA active region 12673 was the site of 4 X-ray class flares and many M-class flares in September 2017.
The evolution of this active region and associated solar flares has been extensively investigated using numerical models \citep{Liu2019, Moraitis2019, Price2019, Inoue2021}. 

\subsection{Case-study of 6 September 2017}

In this review, we present the work of \citet{Guojh2024} and discuss the robustness of the proposed data-driven model. They developed a full data-driven model to study the long-term evolution of this active region and its produced confined X2.2 flare of 6 September 2017 (SOL2017-09-06T08:57), which occurred a few hours before the large X 9.3 flare in AR NOAA 12673. 
{ The question of the confinement of this flare was debated in  \citet{Liu2018}. 
At 09:48 UT, two bright points are visible in the inner part of LASCO/C2 and develop as faint coronal flows. However, several points show that these coronal flows do not correspond to a CME expelled from the active region: the non-lateral extension of flare ribbons and the non-visibility of dimmings, and finally, the velocity estimated from the duration between the flare onset and observed CME is larger than that of this coronal flow.  On the contrary, the second flare X9.3 presents the former characteristics and is very eruptive \citep{Jiang2018}.}

In this data-driven modelling, the initial magnetic field is the potential field model, and the bottom boundary is provided by the time series of the observed vector magnetograms during one day and derived DAVE4VM velocity fields. { 300 vector magnetograms have been used for this study.} As a result, the bottom boundaries of the simulation are synchronised with the observations at every computation time step. { In the data-constrain models used 
for the case studies of 28 October 2021 and 18 August 2022, 
the initial magnetic field is provided by the NLFFF extrapolation of one vector magnetogram and the incorporation of flux ropes;  the driven duration is within 2 hours until the eruption.  In the full data-driven model proposed in \citet{Guojh2024}  the flux rope is not incorporated but formed through the long-term evolution of the active region.}  
The final state of the time-dependent magnetofrictional (TMF) modelling further serves as the initial prerequisite for the thermodynamic MHD simulation, enabling us to investigate the formation and eruption of the observed flux rope.


\begin{figure}[ht]
\centering
\includegraphics[width=0.99\textwidth]{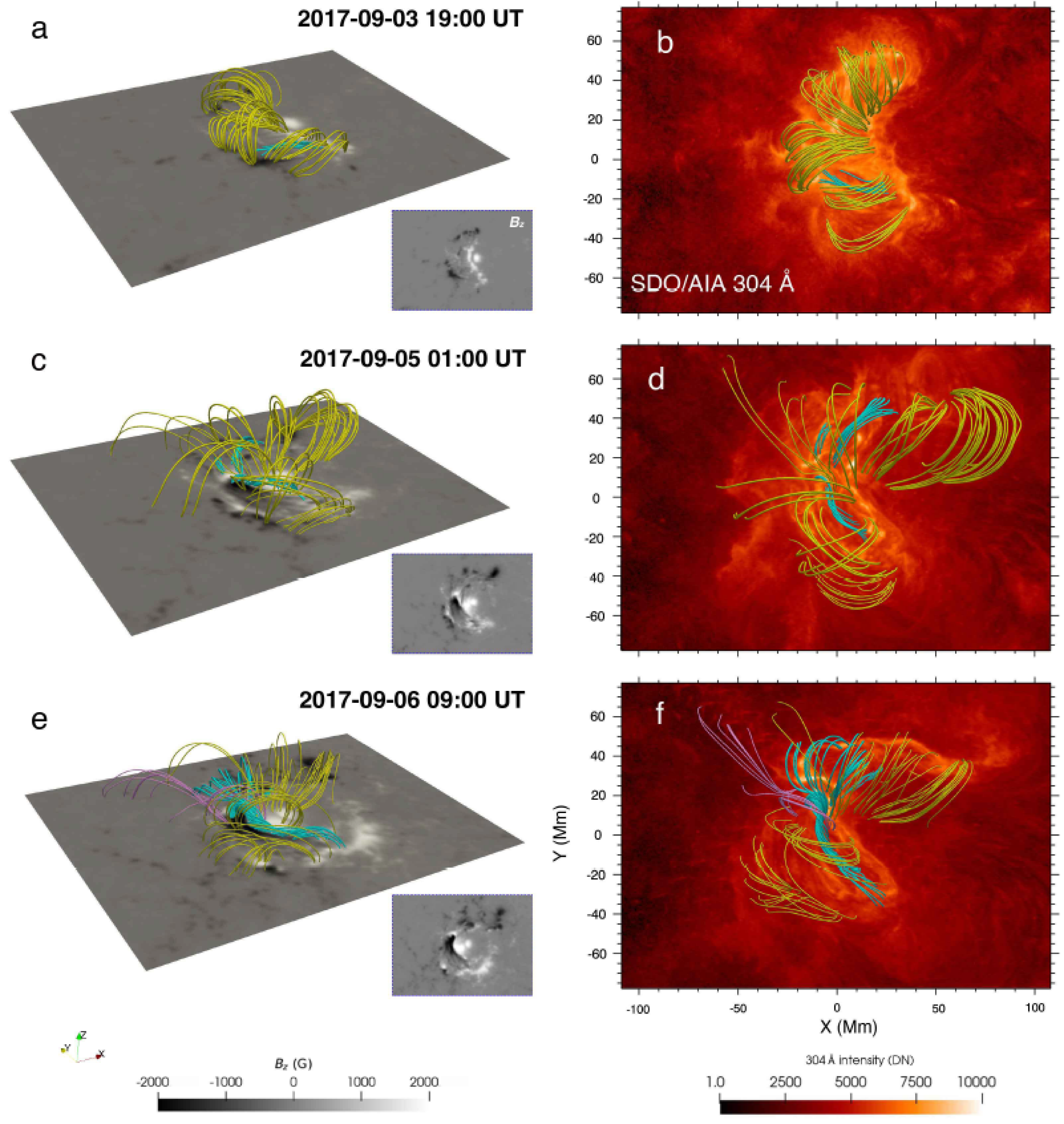}
\caption{Left panels (a,c,e):  Evolution of the flux rope with the magneto frictional method for three time steps and the corresponding magnetograms. Right panels (b,d,f) Comparison with AIA 304 \AA\ images (adapted from \citet{Guojh2024}). 
}\label{FR_AIA}
\end{figure}

\begin{figure}[ht]
\centering
\includegraphics[width=0.99\textwidth]{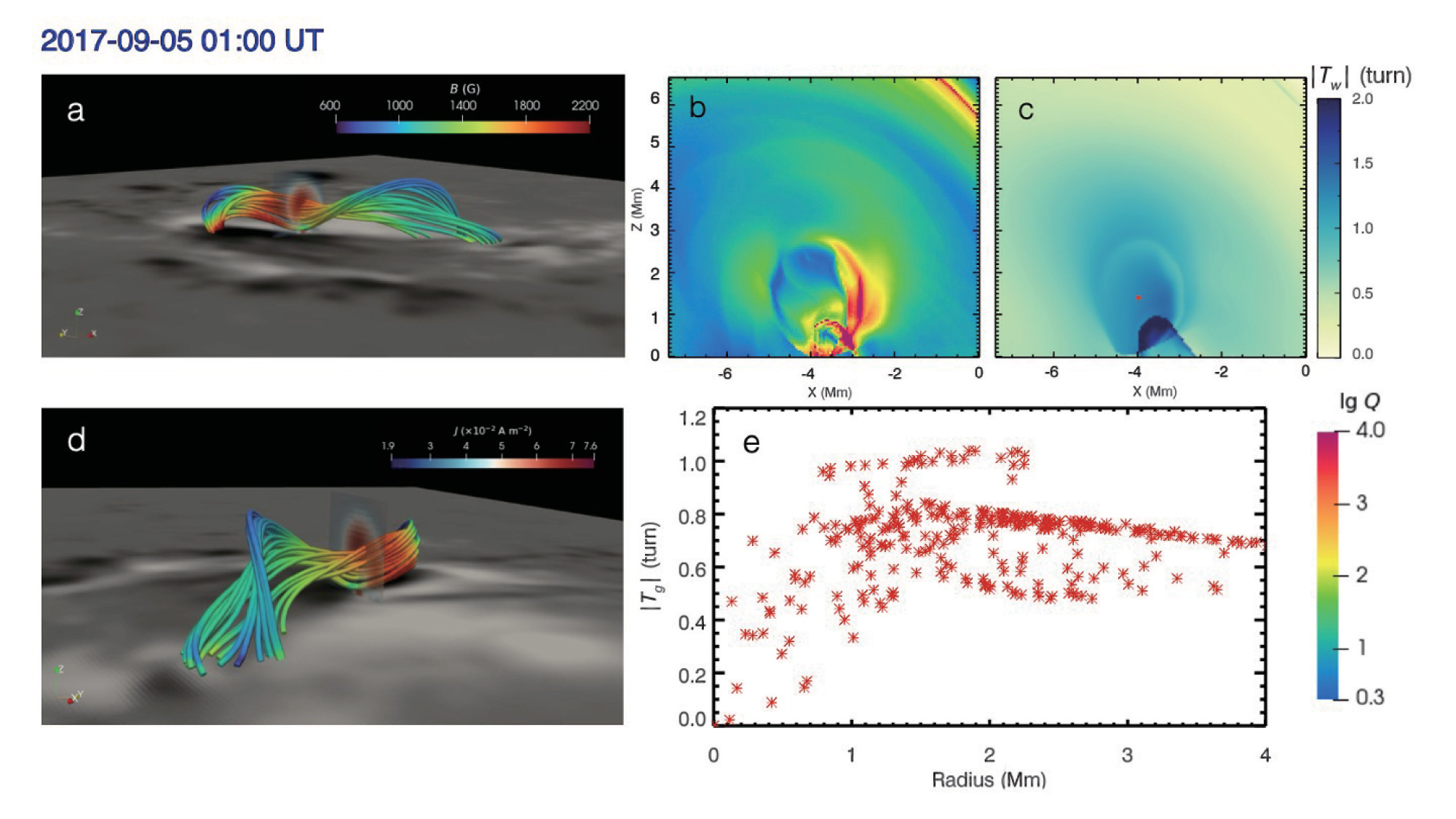}
\caption{Twisted flux rope and magnetic topological structures obtained in a data-driven MHD simulation at 01:00 UT on  5 September 2017.
Panels (a) and (d) illustrate two views of the formed flux rope, where the semi-transparent vertical slices across the flux rope represent the electric-current channel. 
Panels (b), (c) and (e) present the squashing factor ($Q$),  the twist ($T_{w}$) maps in the same planes as panels (a) and (d). Panel (e) shows the $T_{g}$ numbers of selected field lines along the distance from the red dot in panel (c) (adapted from \citet{Guojh2024}). 
}
\label{FR_QSL}
\end{figure}

\begin{figure}[ht]
\centering
\includegraphics[width=0.99\textwidth]{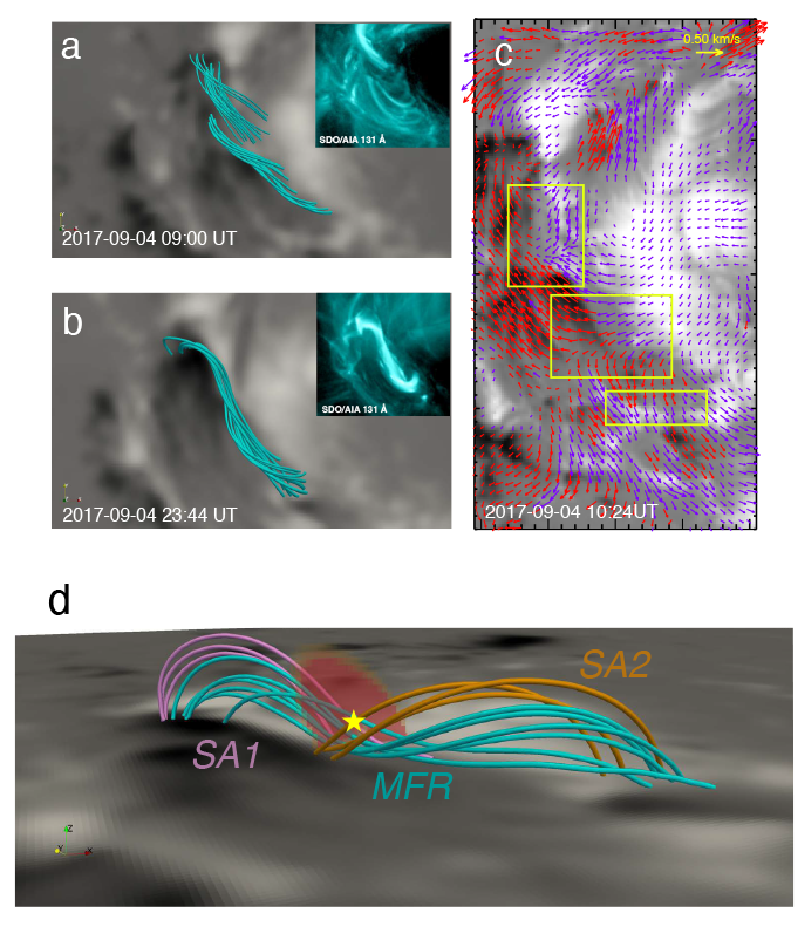}
\caption{Formation of a flux rope simulated by the data-driven model. Panels (a) and (b) display the sheared arcades and flux ropes, respectively. The inserts show the
SDO/AIA 131 Å images at the same time. Panel (c) displays the photospheric horizontal velocity fields.
Panel (d) illustrates the reconnection configuration of the flux rope. The field lines traced from the current sheet depicted by the high $J/B$ region can be divided into two groups of sheared arcades before reconnection (SA1 and SA2) and the newly reconnected flux rope (adapted from \citet{Guojh2024}).
}
\label{FR_formation}
\end{figure}

They employ a data-driven technique to study the flux rope's entire process, from ts birth to its  eruption. The initial magnetic field, before the eruption, is a potential field, and the subsequent coronal evolution is entirely influenced by the observational HMI magnetograms in the photosphere via the TMF approach. Figure~\ref{FR_AIA} illustrates the evolution of the 3D coronal magnetic fields and the comparisons with the SDO/AIA 304~\AA\ observations. It is found that the flux rope formed consistently by coupling the TMF approach with the thermodynamic MHD model \citep{Guojh2024}. The magnetic topology properties of the simulated flux rope are shown in Figure~\ref{FR_QSL}, from which one can see that the twist numbers of certain flux-rope field lines are more significant than one and are recognised as the quasi-circular QSLs. Figure~\ref{FR_formation} reveals the transformation from sheared arcades (Figure~\ref{FR_formation}a) to the flux rope that is well comparable to the observed hot channel in SDO/AIA 131 \AA\ wavelength (Figure~\ref{FR_formation}b), which is formed due to flux cancellation 
driven by collisional shearing motions (Figure~\ref{FR_formation}c). Additionally, it is also found there exists a current sheet represented by a high $J/B$ region, where the traced field lines are composed of two groups of sheared arcades and the central twisted flux rope, as shown in Figure~\ref{FR_formation}d. This data-driven model validates the effectiveness of collisional shearing motions in forming flux ropes in complicated active regions \citep{Chintzoglou2019}.

\subsection{Comparison with other data-driven models}

With the development of observational instruments and numerical techniques, the data-driven MHD simulations have become one novel and advanced method to reconstruct the time-evolving solar corona and, therefore, unveil the underlying physics behind observations. The treatments of governing equations and boundaries are different in various data-driven models. Regarding the governing equation, the models can be divided into TMF model \citep{Cheung2012, Cheung2015, Pomoell2019}, zero-$\beta$ \citep{Guo2019, Guoy2021, Zhong2021, Zhong2023, Kaneko2021}, isothermal \citep{Jiang2016, Jiang2018, Jiang2023b}, thermodynamic MHD \citep{Guojha2023, Guojhb2023}, and the hybrid models \citep{Guojh2024, Afanasyev2023, Daei2023}. Among them, the evolution of the TMF model is assumed to be quasi-static such that it is pretty suitable to reproduce the long-term evolution of active regions and the flux rope formation at a fast speed. The zero-$\beta$ and isothermal MHD models omit the thermodynamic evolution of the plasma while capturing the rapid evolution of 3D coronal magnetic fields. The thermodynamic data-driven MHD model is more advanced and is capable of retrieving the evolution of magnetic topology evolution and thermodynamics, as shown in \citet{Guojha2023}. The data-driven boundaries can be classified into the $B$ or $v-B$ \citep{Jiang2016, Guo2019}, $v$ \citep{Jiang2021, WangX2023, Kaneko2021} and $E$ \citep{Cheung2012, Pomoell2019} driven boundaries. 
{Each of these boundary conditions has its own advantages in physics and numerical schemes. For example, $B$-driven boundary can reproduce the evolution of vector magnetic fields in observations by replacing a time series of observed magnetograms. However, this option will also introduce numerical magnetic-field divergence induced by the driven boundary. In contrast, $E$-driven and $V$-driven boundaries can better address such numerical issues, although they strongly rely on the inversion method: deriving the velocity fields or electric fields capable of retrieving the evolution of magnetic fields in observations. To this end, several numerical approaches have been proposed, such as the well-known DAVE4VM method \citep{Schuck2008} for deriving photospheric flows, and the PDFI\_SS method for deriving both inductive and non-inductive electric fields \citep{Fisher2020}. \citet{Welsh2007} discussed comparisons of different inversion techniques for photospheric flows. \citet{Wangxy2023b} compared the effects of two types of derived photospheric flows with data-driven MHD simulations. Notably, $E$-driven and $V$-driven boundaries can be effectively coupled with the advanced CT method, ensuring that the divergence of magnetic fields introduced during numerical computation is controlled to the magnitude of machine precision.} \citet{Toriumi2020} compared the ability of the well-known data-driven models to reproduce the solar eruption in a flux emergence simulation. The differences induced by adopted physical models and data-driven boundaries are worth doing in future works.

As previously mentioned, several numerical MHD models have been used to investigate the evolution of the active region (AR NOAA 12673). The associated solar flares, particularly the second X-ray flare, were eruptive
\citep{Liu2019,Moraitis2019,Price2019,Inoue2021}. Therefore, to evaluate the usefulness of the novel
data-driven model \citep{Guojh2024} to reproduce the observations concerning these long lists of models, one must conduct a comparison between the new simulation
results and previous results. The comparison can be conducted from various aspects, including the typical magnetic topology, magnetic relative helicity and energy budgets. Firstly, the results of \citet{Guojh2024} exhibit similar trends and magnitudes in magnetic helicity and energy budgets to the TMF simulation carried out by \citet{Price2019} for the same active region. A similarity in the ratio between the current-carrying helicity and total helicity by September 2017 is noted, approximately at 0.15, though the ratio between magnetic free energy to total magnetic energy in \citet{Guojh2024} is slightly higher. This discrepancy could be attributed to different initial conditions (such as the starting time of the potential field extrapolation) and the driven boundary conditions ($E$ or $v-B$). Regarding the magnetic topology, many papers focusing on modelling this active region identify two null-point reconnection sites at the onset of eruption \citep{Mitra2018, Price2019, Bamba2019, Inoue2021, Daei2023}. This consistency with previous findings validates the robustness of the data-driven model of \citet{Guojh2024} in reproducing observed solar eruptions.

The hybrid model combined with TMF and thermodynamic MHD modellings in \citet{Guojh2024} presents several advances, enabling it to capture both the long-term buildup and subsequent drastic release of magnetic energy with a rapid computation speed. However, this operation could trigger a numerical eruption. On the one hand, the system may undergo a non-smooth transition when switching from the TMF model to the MHD model, such that the selection of the switching time is crucial for the trigger of eruptions, as demonstrated in \citet{Daei2023}. On the other hand, the final magnetic-field state in the TMF model generally cannot perfectly satisfy the force-free condition. As a result, the nontrivial residual Lorentz force may lead to the ascent of the flux rope formed in the TMF model \citet{Afanasyev2023}. Therefore, to study the initiation process of a CME, such as the slow-rise phase and the trigger \citep{Xing2024}, the MHD model going through the entire process from formation to eruption of a CME flux rope is more suitable, as done in \citet{Jiang2023b}.

\section{Confined/eruptive event: conditions}

Theories on eruption mechanisms have also advanced in recent years. The existing models can be divided into two types.
The first type is based on 
magnetic reconnection occurring at high-lying null point \citep{Kusano2012}. 
The reconnection of the overlying magnetic field lines leads to a breakout \citep{Antiochos1999}.
When the magnetic strength is too strong, the flux rope is trapped in a magnetic cage and cannot erupt \citep{2018Natur.554..211A}. The orientation of the magnetic field at the top of the flux rope with the environment can lead to confined eruptive eruptions depending on the parallel or anti-parallel lines \citep{Zuccarello2017}.
A rotation of the flux rope could lead to such a configuration and finally to a confined eruption \citep{Zhou2019, Jiang2023}. 
The reconnection may also occur below the flux rope, corresponding to the tether-cutting mechanism model \citep{Moore1980}. In this model, new flux can be injected, providing an upward Lorentz force to the erupting structure \citep{Moore2001}. 

The second type is based on ideal
magnetohydrodynamic (MHD) instabilities, e.g., the
helical kink instability \citep{Torok2004} or the torus instability \citep{Kliem2006, Aulanier2010}. The torus instability can trigger a flux rope eruption \citep{Guo2019}.

The torus instability is mainly dominated by the hoop force $F_{H}$ and the strapping force $F_{s}$. The threshold of the eruptive event is given by the decay index of the background strapping fields (n$>$1.5) but can be less or more  \citep{Demoulin2010, Zuccarello2015, Zuccarello2016}, for the following reasons. 
On the one hand, the threshold of the decay index is related to the aspect ratio of the flux rope. For example, \citet{Demoulin2010} found that the threshold of the decay index decreases to a value of 1.1 for the thick current. On the other hand, other than the strapping force induced by the poloidal magnetic fields, the toroidal-field tension force \citep{Myers2015, Myers2016}, and the force resulting from the non-axisymmetry of the flux rope can also suppress the rising of the flux rope \citep{Zhong2021}. 

The previous sections have shown that thermodynamic magnetohydrodynamic simulations with refined treatments on
the active region evolution and energy transfer have been greatly developed. These enable us to better understand the trigger of flux rope eruption. In such simulations, the magnetic field and coronal plasma evolve fully self-consistently. The simulations provide synthetic images that can be compared with 
remote-sensing observations of different spacecraft and the values of physical parameters in 3D. Therefore, the causes of eruption or confinement can be described by the involved forces \citep{Chen2023, Wang2023, Guojh2024}.

\begin{figure}[ht]
\centering
\includegraphics[width=0.99\textwidth]{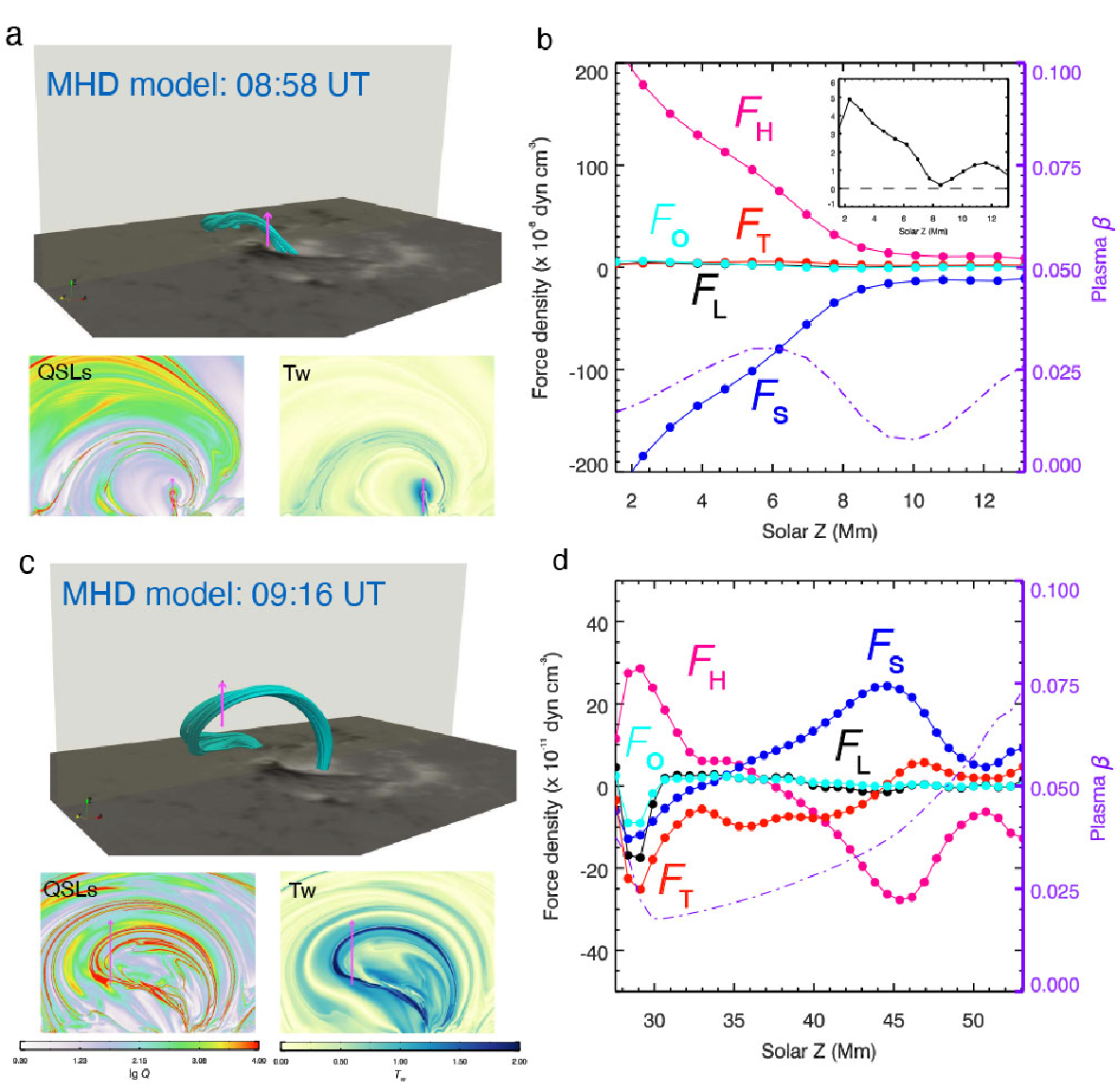}
\caption{
Lorentz force components acting upon the flux rope. Panels (a) and (c) show the 3D illustrations of the magnetic configuration and topology ($Q$ and $T_{w}$ distributions) at the initiation and the end of the eruption. Panels (b) and (d) show the distributions of the Lorentz force components, such as the net Lorentz force ($F_{L}$, black line), hoop force ($F_{H}$, pink line), strapping force ($F_{S}$, blue line), tension force ($F_{T}$, red line) and residual force ($F_{O}$, cyan line). The {\bf purple dash-dotted line} represents the plasma $\beta$
(adapted from \citet{Guojh2024}).
}
\label{Lorentz}
\end{figure}

\begin{figure}[ht]
\centering
\includegraphics[width=0.99\textwidth]{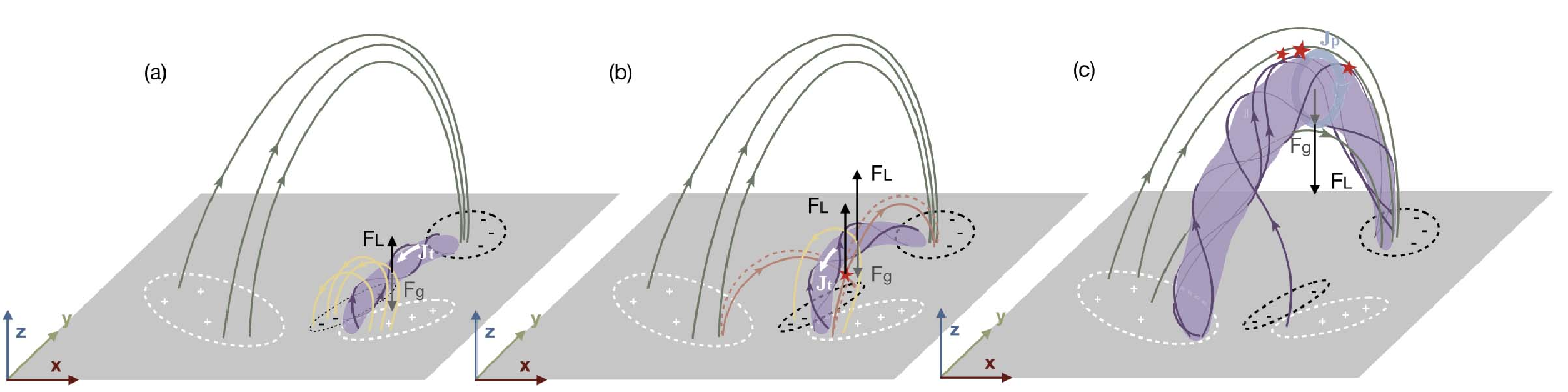}
\caption{Sketch of the confined eruption describing the initiation and confinement of the solar eruption. The yellow and dark green lines represent the background toroidal and poloidal magnetic fields. The purple tubes show the flux ropes. The red pentagrams indicate the potential reconnection regions (adapted from \citet{Wang2023}).}
\label{toroidal}
\end{figure}

\citet{Guojh2024} unveils the  nature of the confined X2.2 eruption on 6 September 2017  with a data-driven simulation. They found that the rotation of the flux rope can lead to the transformation of the overlying magnetic fields from the poloidal to the toroidal direction, thereby increasing the toroidal tension force while decreasing the poloidal strapping force. This is also mentioned in \citet{Zhong2023}.  Figure~\ref{Lorentz} exhibits the Lorentz force involved in the flux rope ejections, including the net Lorentz force ($F_{L}$), the hoop force ($F_{H}$) and strapping force ($F_{S}$) and the tension force {\bf ($F_{T}$)}. As illustrated in Figure~\ref{Lorentz}d, the rising of the flux rope is dominantly constrained by the tension force. More intriguingly, \citet{Zhang2024} studied the constraints of a failed filament eruption associated with the large-angle rotation. They found that the direction of the strapping force becomes upward after the flux-rope rotation that is larger than 90$^{\circ}$, meaning that the poloidal-field strapping force cannot serve as the constraints for the eruption events with the large-angle rotation. This also self-consistently explains why so many failed filament eruptions with the large-angle rotation are torus-unstable \citep{Zhou2019}. Figure~\ref{toroidal} provided a sketch to show a failed eruption constrained by the tension force induced by overlying toroidal magnetic fields \citep{Wang2023}.


It should be noted that the effects of the tension force in constraining solar eruptions have also previously been found in observations \citep{Joshi2022}. They computed the three components of the magnetic fields directly derived from the PFSS model and demonstrated that the tension force was more significant above the active region and led to a confined flare \citep{Joshi2022}.

\section{Conclusion}\label{sec13}

In this paper, we have reviewed { a few} advanced numerical MHD simulations concerning solar eruptions. All of them  
include two steps: modelling setup and numerical solving of MHD equations. In the setup stage, the primary input is the 3D magnetic fields as the initial condition, which can be provided by the potential field model or the results reconstructed from the NLFFF extrapolation. The bottom boundaries are constrained or driven by observed vector magnetograms and photospheric flows. 

\begin{enumerate}
    \item If we aim to investigate the accumulation process of magnetic free energy and helicity, the potential field should establish the initial coronal magnetic field in data-driven models. The coronal magnetic fields become sheared and twisted, driven by input data-driven boundaries such as the vector magnetograms and photospheric flows. In addition, { the starting point of the simulation should be a time when the field is close to the potential field,}
      far from the onset time of the eruption.

    \item If we mainly talk about the eruption process rather than studying the established method of the eruptive flux rope, the initial magnetic fields can be directly provided by the NLFFF extrapolation. In this case, the core fields for the eruption, such as flux ropes, could be included in the NLFFF extrapolation, namely, the initial magnetic field condition for the subsequent MHD simulation. The magneto-frictional relaxation combined with the RBSL flux-rope insertion is just used to construct the NLFFF, which is adopted in \citet{Guoy2021, Guojha2023, Guojhb2023}. 

    \item More recently has been developed a global coronal model (COCONUT- \citet{Perri2022,Perri2023} ), where a realistic solar wind reconstructed from observed magnetograms is superposed in the corona \citep{Linan2023, Guojh2024b}. The implementation of the RBSL flux rope model in COCONUT is promising and can be coupled to EUHFORIA simulations to predict space weather events in the Earth's environment \citep{Pomoell2018,Poedts_2018}.
 
\end{enumerate}

These new thermodynamic data-driven MHD simulations allow us to understand eruptions and the forces involved in confined or eruptive events. They follow the actual Sun and open a new domain of research, which leads to better predictions of eruptive phenomena and partition of the release of magnetic energy in the atmosphere.






\bmhead{Acknowledgements}
These results were also obtained in the framework of the projects
C14/19/089  (C1 project Internal Funds KU Leuven), G0B5823N and G002523N (WEAVE)   (FWO-Vlaanderen), 4000134474 (SIDC Data Exploitation, ESA Prodex), and Belspo project B2/191/P1/SWiM.
For the computations, we used the VSC—Flemish Supercomputer Center infrastructure, funded by the Hercules Foundation and the Flemish Government Department EWI. BS wants to thank FWO for supporting her trip to Nagoya for the 7th Asia-Pacific Conference on Plasma Physics from November 12 to 17, where she presented this review.
 We are indebted to Luis Linan for his fruitful comments on this review.

\noindent

\end{document}